# High-Capacity High-Power Thermal Energy Storage Using Solid-Solid Martensitic Transformations

**AUTHORS:** Darin J. Sharar[1*], Asher C. Leff[1,2], Adam A. Wilson[1], Andrew Smith[3]

**AFFILIATIONS**

[1]U.S. Army Research Laboratory, Adelphi, MD 20783, United States

[2]General Technical Services LLC, Wall, NJ 07727, United States

[3]U.S. Naval Academy, Annapolis, MD 21402, United States

**CORRESPONDING AUTHOR**

*Correspondence and requests for materials should be addressed to:

Darin James Sharar, Ph.D.

U.S. Army Research Laboratory

2800 Powder Mill Rd.

Adelphi, MD 20783

darin.j.sharar.civ@mail.mil

**NOTES TO PUBLISHER**

All figures (supplementary information and main text) should be printed in color for best reading experience.

No copyrighted material has been used.

This work has not been published elsewhere.

**ABSTRACT**

Adding thermal conductivity enhancements to increase thermal power in solid-liquid phase-change thermal energy storage modules compromises volumetric energy density and often times reduces the mass and volume of active phase change material (PCM) by well over half. In this study, a new concept of building thermal energy storage modules using high-conductivity, solid-solid, shape memory alloys is demonstrated to eliminate this trade-off and enable devices that have both high heat transfer rate and high thermal capacity. Nickel titanium, $Ni_{50.28}Ti_{49.36}$, was solution heat treated and characterized using differential scanning calorimetry and Xenon Flash to determine transformation temperature (78˚C), latent heat (183 kJm$^{-3}$), and thermal conductivity in the Austenite and Martensite phases (12.92/12.64 Wm$^{-1}$K$^{-1}$). Four parallel-plate thermal energy storage demonstrators were designed, fabricated, and tested in a thermofluidic test setup. These included a baseline sensible heating module (aluminum), a conventional solid-liquid PCM module (aluminum/1-octadecanol), an all-solid-solid PCM module ($Ni_{50.28}Ti_{49.36}$), and a composite solid-solid/solid-liquid PCM module ($Ni_{50.28}Ti_{49.36}$/1-octadecanol). By using high-conductivity solid-solid PCMs, and eliminating the need for encapsulants and conductivity enhancements, we are able to demonstrate a 1.73-3.38 times improvement in volumetric thermal capacity and a 2.03-3.21 times improvement in power density as compared to the conventional approaches. These experimental results are bolstered by analytical models to explain the observed heat transfer physics and reveal a 5.86 times improvement in thermal time constant. This work demonstrates the ability to build high-capacity and high-power thermal energy storage modules using multifunctional shape memory alloys and opens the door for leap ahead improvement in thermal energy storage performance.

**HIGHLIGHTS**

• First-of-a-kind Nickel Titanium-based thermal energy storage modules were fabricated.

• High-power and -capacity thermal energy storage was demonstrated using Nickel Titanium.

• The maximum power density is 0.848 W/cm$^3$, 2.03-3.21 times higher than standard approaches.

• Module capacity was increased by 1.73-3.38 times.

• Module time constant was improved by 5.86 times.

## NOMENCLATURE

A          Convective heat transfer area, $m^2$

$C_p$      Specific heat, $Jkg^{-1}K^{-1}$

e          Euler number

E          Energy absorbed or discharged, J

FOM        Figure of Merit, $J^2K^{-1}s^{-1}m^{-4}$

h          Heat transfer coefficient, $Wm^{-2}K^{-1}$

k          Thermal conductivity, $Wm^{-1}K^{-1}$

l          Thickness, m

L          Latent heat, $Jkg^{-1}$

m          Mass, kg

ṁ          Mass flowrate, $kgs^{-1}$

PCM        Phase change material

q          Heat, W

SL         Solid-liquid phase change

SMA        Shape Memory Alloy

SS         Solid-solid phase change

t          Time, seconds

T          Temperature, K or ˚C

V          Volume, $m^3$

## GREEK ALPHABET

Δ          Difference in temperature

ρ          Density, $kgm^{-3}$

τ          Thermal time constant, s

## SUBSCRIPTS

eff                  Effective heat transfer coefficient

final                Final steady state temperature

high temp            Temperature above the phase transformation

in                   Inlet fluid temperature

initial              Initial temperature

instantaneous        Measurement at an instant in time

| | |
|---|---|
| latent or L | Latent heating |
| loss | Denoting parasitic energy lost to the environment |
| low temp | Temperature below the phase transformation |
| out | Outlet fluid temperature |
| s | Convective heat transfer area |
| sensible | Sensible heating |
| steady state | Unvarying condition in heating or cooling process |
| t | Value at time (t) |
| t + tf | Time shifted value at time (t+tf) |
| tof | Measurement corrected to account for fluid time of flight |
| transformation | Relating to a phase transformation |
| true | Measurement corrected to account for fluid time of flight and losses |
| tubing | Specifying fluidic tubing and connectors |
| water | Specified property or state of water |

## 1. Introduction

High-power electronics, RF amplifiers, and laser photonic devices are theoretically capable of unlimited on-time, however, high heat fluxes combined with limited maximum device temperatures prevent prolonged operation. Continuous air, liquid, and vapor compression cooling can be scaled-up to handle high thermal loads but the additional size, weight and power (SWaP) required for such scaled-up thermal management systems makes these approaches impractical [1]. Instead, pulse power operation modes coupled with transient thermal management approaches are being adopted to balance high-power needs with platform considerations [2]. One promising approach is the use of thermal energy storage (TES) to passively store and release thermal energy; a summary of physical TES solutions, which can be classified by the method used to store heat, are shown in Figure 1. The combination of TES and pulse power operation lowers the time-averaged thermal load on the primary coolant loop. This, in turn, reduces the overall thermal management power consumption, enables a reduction in the size of ancillary components (fluid reservoirs, radiators, fans, heat exchangers, and pumps), and provides a method for balancing capability and practicality on SWaP-constrained platforms.

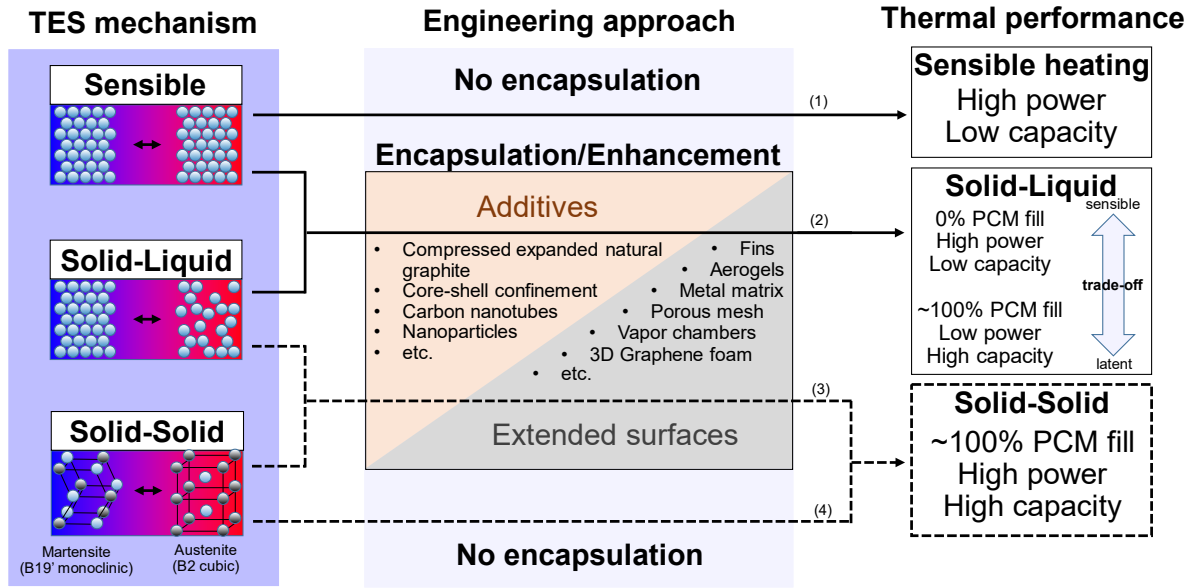

**Figure 1:** Overview of physical thermal energy storage (TES) material classes (left), engineering approaches for module/device fabrication (middle), and relative thermal performance (right).

Among available approaches, thermal energy storage using organic solid-to-liquid phase change materials (SL-PCMs) has gained considerable attention owing to their cost effectiveness, suitable melting temperatures for electronic and photonic cooling, and near-isothermal phase transitions that temporarily result in a very high thermal capacitance. The latter allows thermal energy storage at a preferred temperature with minimal material weight and volume [3]. For example, to store 225 kJ of thermal energy roughly 1 kg of 1-Octadecanol (225 kJkg$^{-1}$) or Paraffin (200 kJkg$^{-1}$) would need to be melted while the same mass of aluminum (0.9 kJkg$^{-1}$K$^{-1}$) would need to be heated roughly 250˚C. This PCM storage density benefit has led to commercial adoption in low-frequency transient applications aimed to regulate body and building temperatures [4] [5] [6], modulate discrepancies between peak solar thermal energy absorption and peak demand [7] [8] [9], and peak shave HVAC systems [10]. However, transition to high heat flux, high power, fast-transient applications has been hindered by integration and performance concerns. Solid-liquid PCMs require engineering measures in the form of specially-engineered metallic fin structures and additives to provide mechanical support, accommodate large solid-to-liquid volumetric changes (10-15% typical), prevent liquid phase PCM leakage, and enhance the inadequate PCM thermal conductivity (typically 0.1 to 1 Wm$^{-1}$K$^{-1}$ [11]); in particular, poor thermal conductivity results in large time constants and inhibits fast, high power operation.

The need for encapsulation and the goal of increasing power by adding high thermal conductivity sensible heating materials has come at the expense of reduced module energy capacity [12] [13], as described schematically in Figure 1. In many cases, this reduces the mass and volume of active PCM material by well over half. For example, Medrano et al. [14] evaluated five commercial heat exchangers for use as PCM thermal storage systems and reported PCM to total weight ratios ranging from 0.024 to 0.579. They performed charging and discharging studies using water and found that a double pipe heat exchanger with commercial paraffin RT35 (0.20 $Wm^{-1}K^{-1}$) embedded in a graphite matrix (PCM weight ratio of 0.24) provided the highest power (0.12 $Wcm^{-3}$). Yang et al. [15] employed copper foam and fins and reported 1/3 less time for complete paraffin melting (0.06 $Wcm^{-3}$), but at the cost of a 0.51 PCM to total weight ratio. Mallow et al. [16] used compressed expanded natural graphite (CENG) and paraffin composites to improve the effective thermal conductivity from 0.28 to 10.1 $Wm^{-1}K^{-1}$. The consequence was a 32% reduction in the effective latent heat, from 163.4 to 111.1 $kJkg^{-1}$. Others have taken a more subtle approach to increasing thermal conductivity in solid-liquid PCMs by using lower additive weight percentages, but with less dramatic improvements in thermal conductivity [17] [18] [19] [20]. As an example, a recent study by Al-Ahmed et al. [21] produced Octadecanol/MWCNT (5 wt%) and Octadecanol/OD-g-MWCNT (5 wt%) composite PCMs with enthalpy values very close to pure Octadecanol, but with thermal conductivities of only 0.56 to 0.58 $Wm^{-1}K^{-1}$ (baseline 0.25 $Wm^{-1}K^{-1}$). Even using Additive Manufacturing (AM) to fabricate composite structures, which enables unparalleled design flexibility and optimization, necessarily requires sacrificing PCM volume for conductive fins. As shown by Moon et al. [22], a PCM volume fraction of 0.85 and an estimated paraffin to AM aluminum silicon alloy (AlSi10Mg) weight ratio of 0.64 was used to improve TES power density (0.58 $Wcm^{-3}$); notably, the authors report that a power density of 0.58 $Wcm^{-3}$ represents a 4X improvement over conventional designs and the largest published liquid-to-TES heat exchanger value to date. More recently, Iradukunda et al. [23] used topology optimization schemes and Direct Metal Laser Sintering (DMLS) of AlSi10Mg to design and fabricate optimal fin geometries for D-Sorbitol (0.59 $Wm^{-1}K^{-1}$) solid-liquid electronic heat sinks. As a baseline, they modeled straight fins and reported a trade-off between melt speed and melt duration with PCM volume fractions from 0.6 to 1.0. Ultimately, they settled on a PCM volume fraction of 0.70 in absence of power, temperature, and storage requirements. Regardless of the approach, the standard method of swapping PCM volume for additive and/or fin volume to improve thermal conductivity remains a key technical barrier for high-power, compact, thermal energy storage.

Sharar et al. [24] [25] recently identified the use of reversible solid-solid Martensitic transformations in NiTi shape memory alloys (SMAs) as high performance thermal energy storage materials. In addition to high volumetric latent heat, approaching or often exceeding that of standard organic PCMs (225 MJm$^{-3}$ [24]), NiTi alloys offer two-orders-of-magnitude higher thermal conductivity, approaching 28 Wm$^{-1}$K$^{-1}$ [26], excellent corrosion resistance [27], high strength and ductility [28], and good formability via traditional thermomechanical processing, machining, and manufacturing [29]. This unique combination of physical properties presents an opportunity to build multifunctional phase change energy storage, heat transfer, and structural media entirely out of robust metallic NiTi alloys. In contrast to the unavoidable capacity/conductivity tradeoff when designing with solid-liquid PCMs, this promises both improved latent energy storage density and fast transient response. In this article, we fabricate and test two first-of-their-kind NiTi TES designs as a proof-of-concept, including one entirely solid-state and one combined solid-state/1-octadecanol module. The NiTi-based modules, including two baselines for comparison, are fabricated and experimentally characterized in a water flow loop from 15 to 80˚C to simulate transient performance in a high power electronic system and provide side-by-side comparison with existing approaches. Analytical models are used to explain our convective heat transfer, module thermal capacity, thermal power, and thermal time constant results, and extrapolate to additional use cases.

## 2. Material Selection, Preparation, and Characterization

Ni$_{50.28}$Ti$_{49.36}$ was purchased from EdgeTech, LLC. to be used as the solid-solid PCM in this study. A full elemental description of this material can be found in the Supplementary Information, along with characterization results (Figure S1) for two additional compositions not chosen for prototyping (Ni$_{50.31}$Ti$_{49.32}$, and Ni$_{50.41}$Ti$_{49.20}$). The NiTi material was fabricated in a rolled sheet form and was solution heat treated in-house at 850°C for 10 minutes then water quenched to promote recrystallization and grain growth [24]. 1-Octadecanol was selected as the SL-PCM because it is inexpensive, widely available, thermally stable, and does not harbor environmental, personal, or compatibility risks [30]. Aluminum-6061 was used as the material for the sensible heating module and as the encapsulant/fin material for the SL module. It should be noted that there is a growing list of solid-liquid metallic [31] and salt hydrate materials [32] that exhibit high volumetric latent heat and high thermal conductivity [3]. However, these materials were not considered for the current study because they are known to exhibit extreme undercooling [33] [34], incongruent melting and phase separation [35] [36], and incompatibility with standard encapsulation materials such as Aluminum [3] [37].

Differential Scanning Calorimetry (DSC) was carried out for all materials using a Perkin-Elmer 8500 DSC to determine latent heat and transformation temperature, along with baseline characteristics for aluminum. Samples were stabilized for 5 minutes to within $\pm 0.1˚C$ of the set-point start temperature of $0˚C$. After stabilizing, the samples were ramped at a rate of $10˚Cmin^{-1}$ to a maximum temperature of $100˚C$, and then cooled at a rate of $10˚Cmin^{-1}$ back to $0˚C$. DSC results for the as-received and heat treated $Ni_{50.28}Ti_{49.36}$ material are shown in Figure 2a. In both cases, the endothermic transformation occurred near $78˚C$ and the reverse transformation occurred at $38˚C$. The latent heat was shown to increase by 230%, from 12 to 28 $Jg^{-1}$ (183.1 $MJm^{-3}$) for the solution annealed material. As shown by the DSC results in Figure 2b, 1-octadecanol has an endothermic transformation near $60˚C$, reverse transformation near $46˚C$, and a mass-specific latent heat of 225 $kJkg^{-1}$ (182.25 $MJm^{-3}$). Unsurprisingly, Aluminum revealed no transformation in the temperature range tested but serves as a useful data point for calibration and comparison.

Thermal conductivity values for 1-octadecanol (0.25 to 0.15 $Wm^{-1}K^{-1}$) and aluminum 6061 (205 $Wm^{-1}K^{-1}$) were taken from the literature and are listed in Figure 2. The thermal diffusivity of the as-received and solution heat treated $Ni_{50.28}Ti_{49.36}$ material was measured using a TA Instruments DXF 200 high-speed Xenon-pulse delivery source and solid-state PIN detector. The NiTi thermal conductivity at a temperature of $0˚C$ (Martensite phase 12.64 $Wm^{-1}K^{-1}$) and $100˚C$ (Austenite phase 12.92 $Wm^{-1}K^{-1}$) are listed in Figure 2 and were calculated from the measured diffusivity values assuming a density of 6450 $kgm^{-3}$ and specific heat of 469 $Jkg^{-1}K^{-1}$ [26].

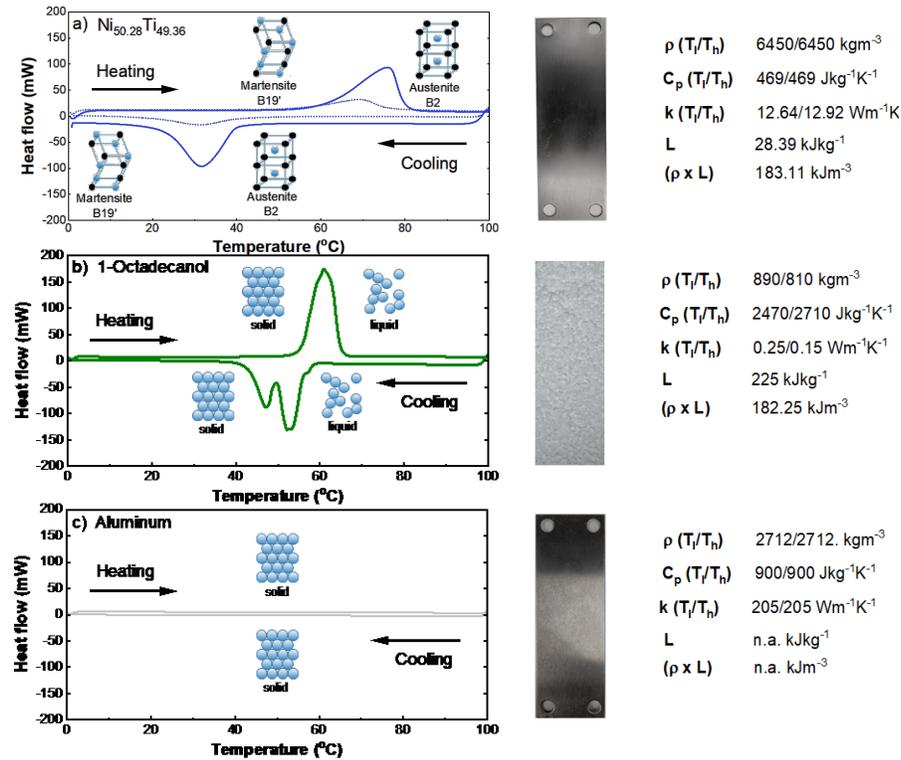

**Figure 2: DSC data (left), photographs (middle), and physical properties (right) of the materials used in the current study: a) Ni$_{50.28}$Ti (dashed line is the as-received material and solid line is the heat treated material), b) 1-Octadecanol and c) aluminum 6061**

## 3. Energy storage module design, fabrication, and testing

The energy storage modules tested in this study were modeled after parallel-plate heat exchangers frequently used in commercial and domestic hot water applications [38] [14]. This allowed comparison to past studies and enables module fabrication with off-the-shelf aluminum and NiTi sheet material. Using the aforementioned materials, four different parallel plate thermal energy storage modules were considered for this study, as shown schematically in Figure 3: (1) a ten plate aluminum sensible energy storage module, (2) the same aluminum module with 1-octadecanol organic SL-PCM filled in the space between adjacent aluminum plates, (3) a composite latent energy storage module with 1-octadecanol in the space between adjacent solution heat treated Ni$_{50.28}$Ti$_{49.36}$ plates, and (4) a ten plate all solid-state PCM module made out of solution heat treated Ni$_{50.28}$Ti$_{49.36}$. The numbers (1)-(4) above correspond with notations in Figure 1; the first two cases represent standard sensible and SL approaches with encapsulation/fins, while the latter two are proof-of-concept solid-solid NiTi designs.

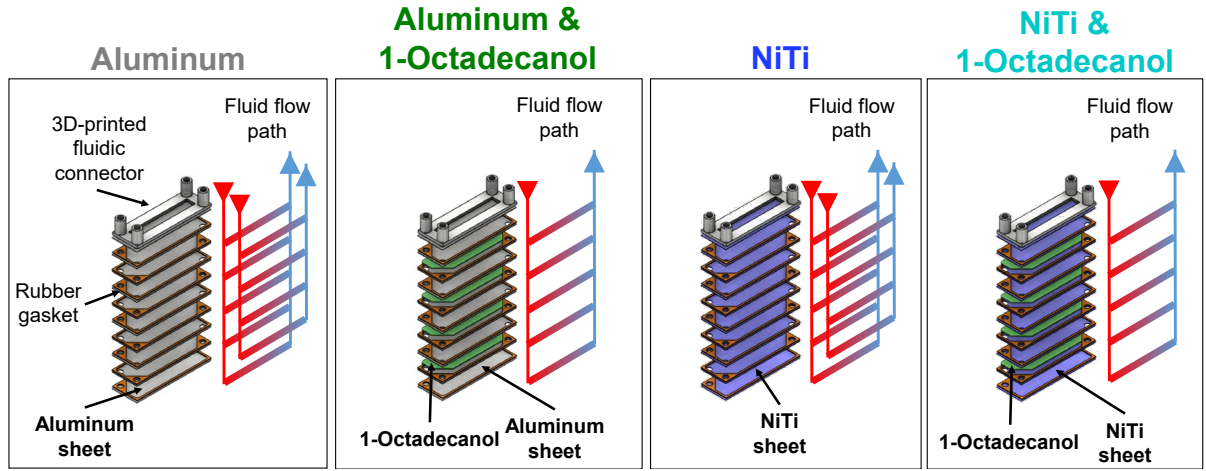

**Figure 3: 3D renderings of the four thermal energy storage designs selected for fabrication and testing. The rubber gasket material is shown in red, the 3D-printed fluidic connector is shown in white, the aluminum material is shown in grey, the 1-octadecanol material is shown in green, and the NiTi is shown in blue.**

In all cases, the Aluminum and NiTi metal sheets were 1 mm thick, 50.8 mm wide, and 152.4 mm long. A metal punch was used to create 9.525mm (3/8") holes in the corners of the metal sheets for fluidic distribution and PCM impregnation, as shown in Figure 2 and schematically in Figure 3. The gaskets were cut from 1.5mm thick gasket rubber and assembled to allow encapsulation of PCM between alternating channels. Fluidic connectors were fabricated using a Stratasys FDM 3D printer. All modules were assembled using a thin coating of Permatex UltraCopper RTV on the gasket material to ensure a robust fluidic seal. The modules containing 1-octadecanol were first fabricated and tested, then placed in a furnace at 90°C (well above the 1-octadecanol melting temperature) for 2 hours to thermally equilibrate. Finally, 41mL (33.2 g) of molten 1-octadecanol was poured into each of the heated TES modules and allowed to cool overnight to room temperature.

The fluidic test setup shown in Figure 4 was developed to test thermal charging and discharging performance in the aforementioned thermal energy storage modules. An Anova recirculating bath was used to chill and pump 15°C water at a fixed mass flowrate of 350 g/min. The flowrate was measured using an Atrato ultrasonic flow meter with a 1.5% absolute reading accuracy and 250:1 turndown ratio (20 mL/min to 5 L/min). 900W of power was supplied to an Omegalux inline heater directly upstream of the TES module to simulate a high power electronic, RF, or laser photonic device. This resulted in heating of the incoming fluid, which in turn promoted transient thermal response in the modules. The mass flux, 350 g/min, was selected to ensure the water was heated to 80°C. Larger mass fluxes would result in lower fluid temperature rise and incomplete

NiTi transformation, which was avoided in these experiments. After 10 minutes (600s), the power to the heater was turned off and, because the fluid bath was still at the desired setpoint of 15˚C, the module was thermally discharged and cooled back to the setpoint temperature. The fluid temperature at the inlet and outlet of the test section was measured by two Special Limits of Error (SLE) Omega Type T thermocouples with ±0.5 ºC accuracy. Three thermocouples were embedded in the center-most fluidic channel (labeled $T_{(1)}$, $T_{(2)}$, and $T_{(3)}$) to monitor stream-wise temperature evolution in the modules. One SLE Type T thermocouple was used to measure the fluid temperature before the inline heater. This experiment represents one complete TES charge and discharge cycle. Experimental uncertainties are detailed in the Supplementary Information.

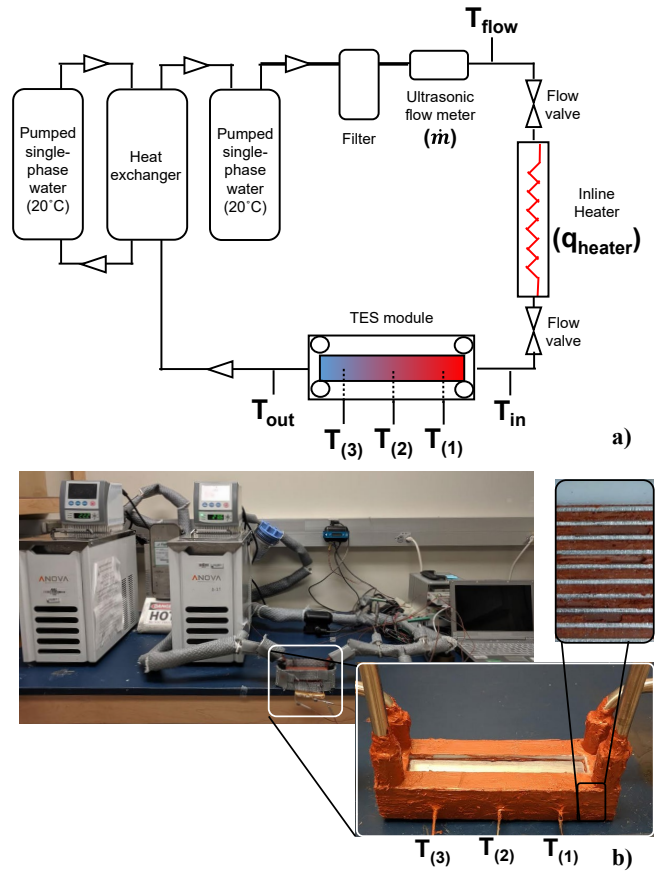

**Figure 4: a) Schematic and b) photograph of the experimental test setup and example TES module used for the current study.**

## 4. Results and Discussion

### 4.1 Module Transient Thermofluidic Response

Figure 5a-h shows DSC and instantaneous thermofluidic testing results for the four material combinations described previously. The DSC (left-most data) and thermofluidic results (right-

most data) share a temperature y-axis, from 0˚C to 95˚C, to allow direct comparison between observed DSC transformation results and module performance during fluidic testing. Furthermore, time scales for all thermofluidic results are kept consistent to allow easy visual comparison of identified points i-ix, which correspond to key inflection points and material transformation temperatures. The observed temperature difference between the inlet and outlet fluid, indicated by the dashed lines, and specific ΔT values corresponding to points i-ix are superimposed.

The result of one fluidic-heating experimental cycle for the aluminum module is shown in Figure 5b. At a time of 0s, both inlet and outlet temperatures are stable at the set-point temperature of 15°C. The fluid begins sensible heating at point (i) once 900W of power is supplied. Near point (ii) at a time of 75s, a peak temperature difference of 16.35°C and apparent peak thermal driving force [14] is observed during the charging event, after which point the rate of heating of the incoming fluid begins to slow and the module and fluid begin to equilibrate. After 600s, point (v), the temperature difference has stabilized to 1.27°C, the heater is turned off, and the module is allowed to cool. During the discharging event, a peak ΔT of 16.40°C is observed at 675s near point (vi), after which the cooling rate slows as the fluid approaches the set-point temperature at point (ix).

It's clear from the side-by-side analysis in Figure 5(a) through (h) that the non-linear temperature responses during transient heating and cooling in the SL, SS, and combined SS/SL storage modules correspond to both the observed peak in the thermal driving force and endo- and exotherms in the DSC data. For example, the SL module fluid ΔT (Figure 5d) shows a clear second peak at point (iii) and a temperature of 60˚C corresponding with the sharp endothermic peak in 1-octadecanol (Figure 5c), resulting in 5.62˚C lower temperature than aluminum. Unsurprisingly, the temperature is nearly the same as aluminum at points (iv) and (v) after the phase transformation is complete. The NiTi module, on the other hand, shows significant deviation from the aluminum and SL modules (even at point iv), but without a clear bimodal peak. As shown in Figure 5e, this is due to the more-gradual phase transformation in NiTi that doesn't conclude until nearly 80˚C; the result is a lower temperature deviation at point (iii) and higher deviation at point (iv) compared to the SL module. Indeed, the combined SS/SL module shares characteristics of both the SS and SL modules, with a defined 1-octadecanol peak at point (iii) and a more gradual NiTi transformation at point (iv).

Similar results can be seen in the discharging responses after point (v). The SL module (Figure 3d) shows a peak during discharging at point (vii) corresponding to the exothermic phase transformation in 1-octadecanol centered at 46˚C. Because the exothermic phase transformation occurs at a lower temperature in NiTi (Figure 3e), the secondary peak is shifted to a later time (point (viii)) corresponding to a temperature of 30 ˚C. The result is a ΔT value of 9.37˚C, as opposed to only 4.53-4.78˚C in the aluminum and SL modules. Finally, the 1-octadecanol and NiTi peaks are superimposed in Figure 5h, resulting in a peak at point (vii) corresponding with the 1-octadecanol phase transformation and sustained performance at point (viii) resulting from the NiTi phase transformation. Eventually, all modules settle back to the setpoint temperature. These results prove that the temperature response during heating and cooling is a direct result of the observed phase transformations, and represent the first experimental demonstration of NiTi in thermal energy storage modules.

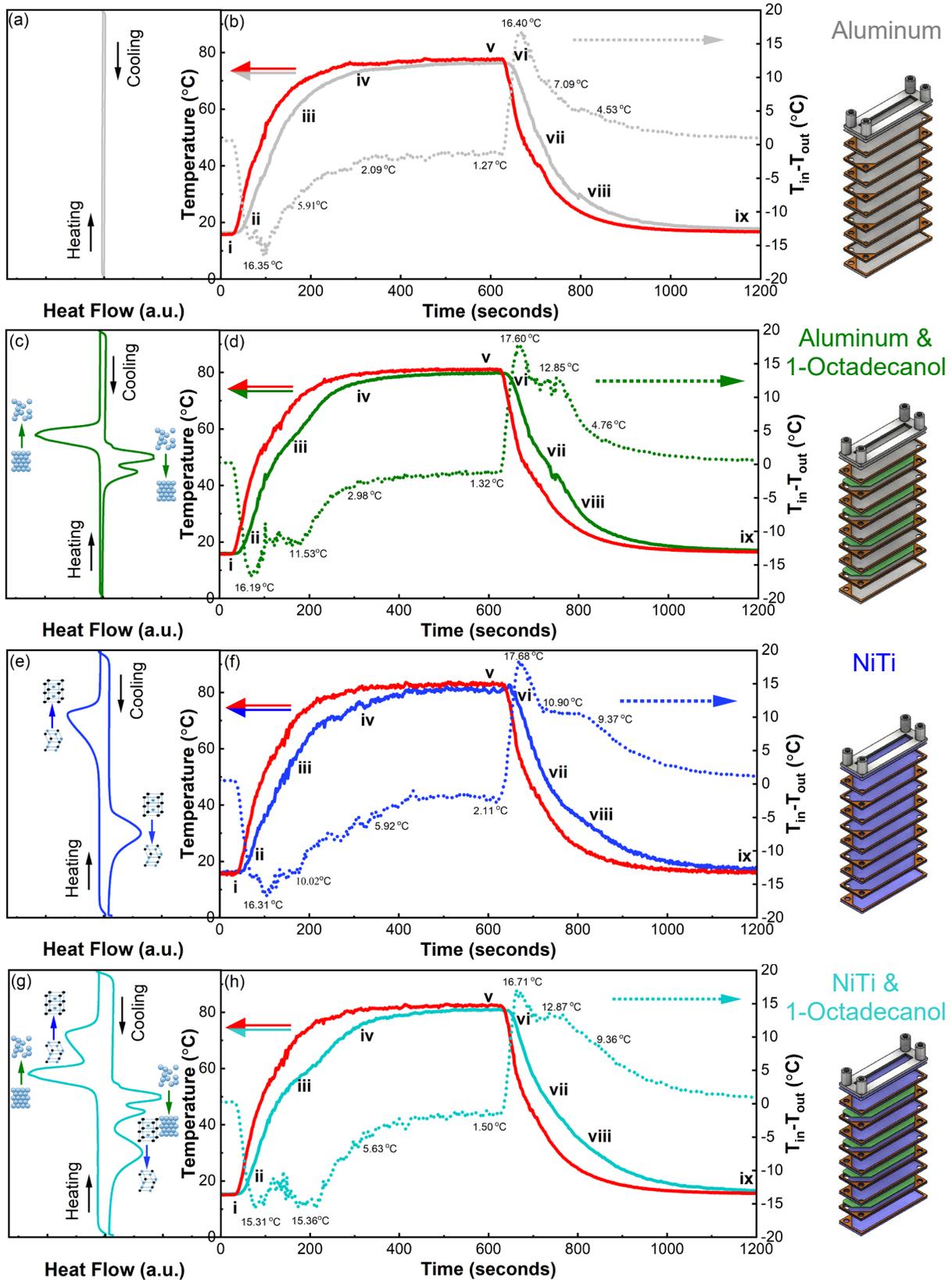

**Figure 5:** DSC results with corresponding phase transformations indicated for (a) Aluminum, (c) 1-Octadecanol, (e) NiTi, and (g) NiTi & 1-Octadecanol. Instantaneous fluid temperature vs time (solid) and superimposed fluid ΔT (dashed) for (b) Aluminum, (d) 1-Octadecanol, (f) NiTi, and (h) NiTi & 1-Octadecanol.

## 4.2 Module Thermal Power

The above analysis is useful for illustrative purposes, however, thermal power and thermal energy are more meaningful quantities. Figure 6 shows the calculated thermal power for all four modules. The dotted lines represent the instantaneous power being absorbed or released by the thermal energy storage modules, which can be determined by the difference in the fluid entering and exiting the structures:

$$q_{instantaneous} = \dot{m}C_{p\ water}(T_{in} - T_{out}) \qquad (1)$$

where $q_{instantaneous}$ is the instantaneous experimental power, $\dot{m}$ is the mass flowrate, $C_{p\ water}$ is the specific heat of water assuming isobaric conditions (assumed to vary linearly between 4.1816 and 4.2157 $Jg^{-1}K^{-1}$ for temperatures from 15 to 100˚C), and $(T_{in} - T_{out})$ is the difference in fluid temperature at the inlet and outlet of the energy storage module, as shown by the dashed lines in Figure 5. Instantaneous power, at a given moment in time, is interesting in application but is not an accurate representation of the power being absorbed or released by the TES module. Instead, a moving reference frame was used, where a voxel of fluid is tracked through the module. The thin solid lines represent the power after accounting for the fluid voxel time-of-flight:

$$q_{tof} = \dot{m}C_{p\ water}(T_{in\ (t)} - T_{out\ (t+tf)}) \qquad (2)$$

where $q_{tof}$ is the adjusted power, (t) represents the reported time, and (tf) is the time it takes the fluid voxel to travel through the reference volume. (tf) was calculated to be between 7 and 15s depending on the specific module and fluidic connectors.

Finally, $q_{loss}$ was subtracted from the time-of-flight power to account for thermal power lost to the environment due to convection and parasitic heating of stainless steel tubing, gasket material, and various connectors. This thermal loss is clear at point (v) in Figure 5 and manifests as a difference between inlet and outlet temperature during steady-state high temperature operation. The thermal loss was calculated to be between 30W and 40W. The true module power, represented by solid bold lines in Figure 6, was calculated:

$$q_{true} = \dot{m}C_{p\ water}(T_{in\ (t)} - T_{out\ (t+tf)}) - q_{loss} \qquad (3)$$

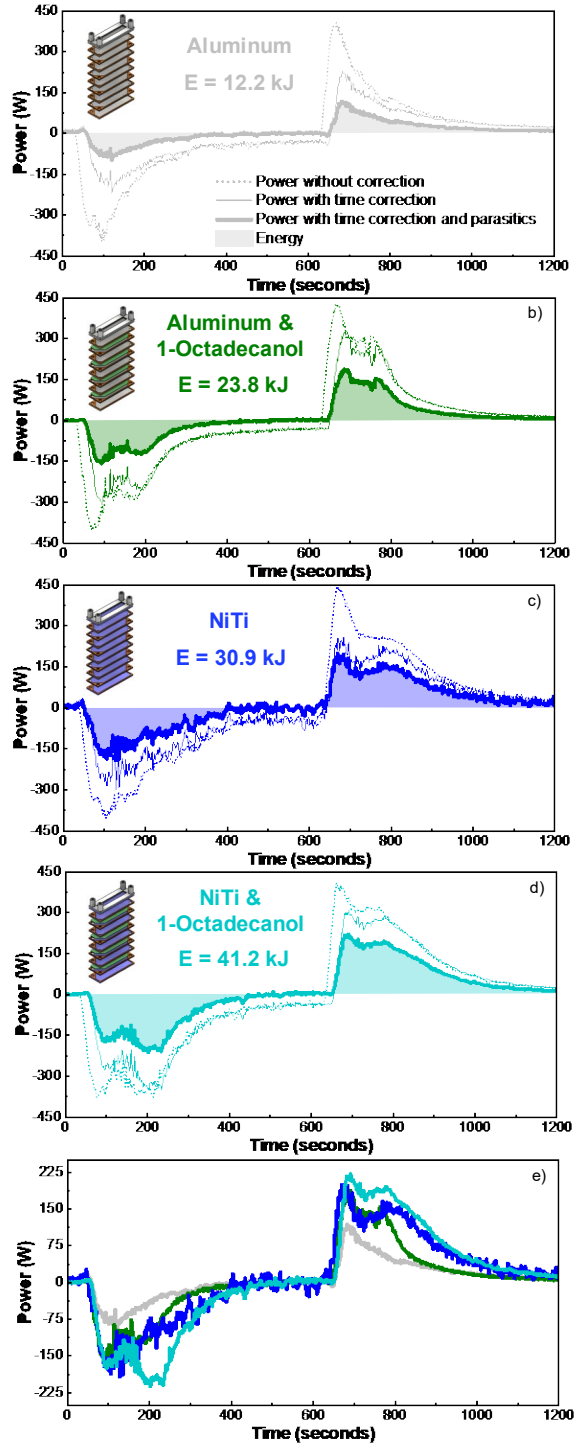

**Figure 6: Raw power data (dotted line), time-of-flight adjusted power (thin solid line), and true power accounting for fluid time-of-flight and parasitic losses (bold solid line) for (a) aluminum module, (b) aluminum & 1-Octadecanol module, (c) NiTi module, and (d) NiTi & 1-Octadecanol module. The shaded regions represent the module energy. (e) shows the true power (Equation 6) for all of the modules.**

These sequential calculation methods result in corresponding reductions in reported thermal power as shown in Figure 6; in the case of aluminum, for example, the calculated peak power during heating reduces from 450W to 98W after correction. The reported peak thermal power for the 1-octadecanol, solid-state NiTi, and NiTi/1-octadecanol module are 161W, 178W, and 218W, respectively; these values are reported in Table 1. This demonstrates the ability to maintain high peak thermal power using phase change heat transfer, particularly NiTi PCMs. The average power was calculated for all modules at an inlet fluid temperature of 78.74˚C, which marks the end of the NiTi transformation and the third RC time constant interval (95% of max). As shown in Table 1, the average power increases from 49.53 W to 158.95 W moving from the Aluminum module to the combined NiTi/1-octadecanol module. Furthermore, for a fixed volume of 187.5 $cm^3$, the reported average volumetric power ranges from 0.264 $Wcm^{-3}$ to 0.848 $Wcm^{-3}$. Past studies report power densities from 0.06 [15] to 0.58 $Wcm^{-3}$ [22]. Our aluminum and 1-octadecanol modules fall within the upper half of these reported values. Notably, our all solid-solid NiTi (0.551 $Wcm^{-3}$) is nearly the same as the recent high-power result by Moon et al. (0.58 $Wcm^{-3}$) [22] and our NiTi/1-octadecanol module (0.848 $Wcm^{-3}$) is 1.46 times higher. The former represents high power and superior manufacturability, eliminating the need for solid-liquid PCMs altogether, and the latter enables ultra-high power density. We attribute the performance improvements to short conductive path lengths in the parallel plate designs and the improved thermal capacity stemming from the use of NiTi.

**Table 1: 3D diagrams, fluidic paths, and salient physical and thermal characteristics of TES modules explored (grey area indicates values calculated based on experimental results in Figures 5 & 6)**

| Description | Al | Al & 1-Octadecanol | NiTi | NiTi & 1-Octadecanol |
|---|---|---|---|---|
| Predicted sensible heat capacity (kJ) | 11.83 | 17.5 | 16.4 | 22.0 |
| Predicted latent energy storage (kJ) SL/SS | 0/0 | 7.5/0 | 0/15.1 | 7.5/15.1 |
| Predicted sensible & latent storage (kJ) | 11.83 | 24.9 | 31.5 | 44.5 |
| Sensible & latent storage (kJ) | 12.1 | 23.8 | 30.9 | 41.2 |
| Peak power (W) | 98.6 | 161.7 | 178.3 | 218.7 |
| Average Power (W) | 49.5 | 100.8 | 103.4 | 158.9 |
| Power Density (W/cm³) | 0.264 | 0.537 | 0.551 | 0.848 |

## 4.3 Module Thermal Capacity

The shaded regions under the power curves in Figure 6a-d represent the total energy absorbed and discharged by the TES modules. The NiTi & 1-octadecanol module had the highest energy storage potential with a value of 41,172 J. This represents a 1.73 and 3.38 times higher energy storage capacity than the standard aluminum/1-octadecanol and aluminum thermal energy storage modules, respectively. Furthermore, the NiTi module offers a caloric benefit of 1.29 to 2.53 times over the aluminum & 1-octadecanol and aluminum modules. The calculated energy capacity values are reported in the shaded region of Table 1.

Sensible heat capacity, latent heat capacity, and combined energy storage values were predicted for a temperature change from 0°C to 100°C, using material properties listed in Figure 2 and Equations S6 and S7 in the Supplementary Information. These results are summarized in Table 1 for a temperature change of 65°C and Figure 7 for the full temperature range. The y-axis intercept on Figure 7a-b represents the latent energy storage of each module; therefore, an assumption has been made that the transformation is nearly instantaneous and always occurs within the specified temperature range. As such, the aluminum module starts at an intercept of zero, while the PCM-based module intercepts correspond with the total latent heat values in Table 1. The slope of the lines represent the specific heat capacity of the respective module. The dashed vertical line in Figure 7 represents a temperature change of 65 ˚C to match the experimental conditions tested. As shown, the experimental results match well with predicted values. The choice to account for fluid time-of-flight and environmental losses are confirmed by the agreement between experimental and analytical energy calculations. Clearly, the area under the raw data curve (dotted line) and time-of-flight shifted line (thin solid line) would over-predict the module power and resultant energy. Moreover, this analysis reveals that the reported improvements in storage capacity are a direct result of the replacement of aluminum with NiTi, which has both higher sensible heat capacity (16.4 kJ vs 11.8 kJ) and added latent heat capacity (15.1 kJ vs 0 kJ), as shown in Table 1. While none of the past or present SL-based designs exceeded a weight ratio of 0.64 while maintaining high power [14] [15] [22] [23], our results demonstrate that solid-solid NiTi PCMs can approach 100% weight and volume ratios since the PCM can also function as the encapsulant structure.

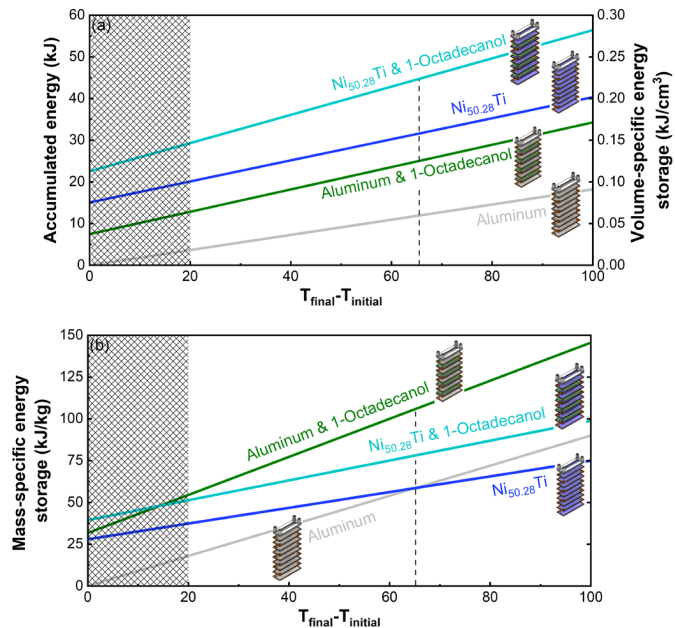

Figure 7: a) Total energy and volume-specific energy storage for the 4 modules fabricated and b) mass-specific energy storage.

Figure 7 provides additional insight into expected performance outside of the conditions tested here. As shown, the NiTi-based modules would always outperform the aluminum-based modules from a total energy and volume-specific energy standpoint for ΔT values from 0 to 100˚C. As shown in Figure 7b, module preference may vary when considering mass-specific energy storage as a dominant metric. However, in a scenario where total energy storage density is a key design factor, which is often the case in SWaP constrained electronic cooling applications, the aluminum-based modules would need to be sized roughly 1.27 to 3.76 times larger to offset the lower volumetric energy storage. As represented by the shaded regions in Figure 7, a temperature change of less than 20°C would result in incomplete NiTi transformation and lower energy absorption. As such, low and near-zero thermal hysteresis SMA materials [39] would need to be considered to realize these performance improvements with low temperature changes, as would be the case for pump diodes and optical gain media [40], which may require forward and reverse transformations in the span of 10°C or less.

Furthermore, higher cycle counts, approaching hundreds and millions of cycles, would be anticipated in many high-power electronic and photonic applications. While such a study is outside of the scope of the current effort, past studies by Chluba et al. [41] and Frenzel et al. [42] have shown at least 10 million transformation cycles in NiTiCu SMAs while maintaining high latent heat (up to 29 Jg$^{-1}$). Transformation temperature and functional properties can also be tailored by altering Ni-Ti balance in binary alloys. As shown in Figure S1 in the <u>Supplemental Information</u>, transformation temperature changed from 15 °C to 80 °C by adjusting the Ni atomic % from 50.41 to 50.28. Others have demonstrated the ability to tune transformation temperature and functional properties by alloying NiTi with Cr, V, and Cu for temperatures between -50 and 100°C [42] and Hf, Pd, Pt, Au, and Zr for high temperature operation up to 500°C [42] [43] [44] [45] [46]. Furthermore, heat treatment to promote grain growth and favorable texture in NiTi materials [26] or the use of high-conductivity Cu-based materials [47], namely CuAlMn and CuAlNi, could offer further improvement in fast-transient applications. So, given appropriate materials synthesis, it is anticipated that SMAs could be applied to a broader range of applications and design points than experimentally validated in this study but would require a rather complicated material optimization to understand property tradeoffs.

### 4.4 Discussion of Material Figure of Merit and Module Thermal Time Constant

Curiously, a marginal improvement in peak power was observed in the NiTi-based modules, even compared to 1-octadecanol which had a similar but slightly lower total storage capacity; intuitively, the higher thermal conductivity (12.22 vs 0.15 Wm$^{-1}$K$^{-1}$) of NiTi should result in higher thermal charge and discharge rates, ie) higher power [48]. To explain this result and develop a more holistic understanding of high-capacity and high-power thermal energy storage, two figures of merit (FOM) are used.

First, we use the material FOM described by Lu [49] as a quantifiable measure of relative PCM performance in high heat flux electronic cooling applications:

$$FOM = \rho \times L \times k_{HT} \qquad (6)$$

Figure 8 shows the calculated figures of merit for the solution heat treated $Ni_{50.28}Ti_{49.36}$ (blue circle) and 1-octadecanol material (green circle) used in our prototypes, as well as organic solid-liquid PCMs, polymeric solid-solid PCMs, solid-liquid salt hydrate PCMs, metallic solid-liquid PCMs, and a range of NiTi-based SMA materials [24]. The calculated FOMs for these materials are $2448.8 \times 10^6$, and $27.3 \times 10^6$ $J^2K^{-1}s^{-1}m^{-4}$, respectively. The high FOM in NiTi materials stems from a comparable volumetric latent heat ($\rho \times L$) and significantly higher thermal conductivity than 1-octadecanol. These results suggest that NiTi can be used as a high-power material replacement for 1-octadecanol without sacrificing volumetric thermal capacity [49]. However, this FOM alone is clearly not adequate for predicting performance in our system.

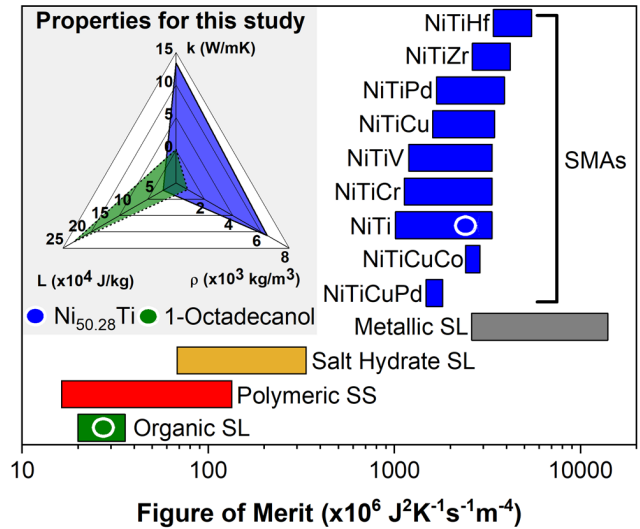

Figure 8: Comparison of TES Figure of Merit for conventional SL and SS PCMs and NiTi-based alloys reported in the literature [17]. The solution heat treated $Ni_{50.28}Ti$ sample characterized herein (blue circle), and SL 1-Octadecanol (green circle) are shown. The inset radar plot shows the measured material properties for the materials tested in this study, and the area represents the calculated Figure of Merit.

Second, we propose modeling the modules as first-order linear time invariant (LTI) systems, or lumped capacity systems. We derived a modified time constant equation using a parallel plate heat transfer coefficient for systems with high Biot numbers and non-negligible conductive resistance [48]:

$$\tau = \frac{\left(\rho C_p V + \frac{L}{T_{final} - T_{initial}}\right)}{\left(\frac{A_s}{\frac{1}{h} + \frac{(l/2)}{3k}}\right)} \qquad (7)$$

where $\rho$ is the density, $C_p$ is the specific heat, $V$ is the volume, $h_L$ is the latent heat, $h_{eff}$ is the effective heat transfer coefficient, and $A_s$ is the convective heat transfer area. In our case, we calculated a heat transfer coefficient of 3950 $Wm^{-2}K^{-1}$ and apparent Biot numbers ranging from 0.01 to 10; Please refer to <u>Supplementary Information</u> and Figure S2 for a description of the heat transfer coefficient calculation [50] and time constant derivation. Using Equation 7, the time constant for the aluminum module is 0.34s, aluminum/1-octadecanol is 5.33s, NiTi is 0.91s, and NiTi/1-octadecanol is 9.65s. The expected temporal response of the modules, provided an instantaneous change in temperature, can be represented by the general equation:

$$\Delta T(t) = (T_{final} - T_{initial})(1 - e^{-t/\tau}) + T_{initial} \qquad (8)$$

where $\Delta T(t)$ is fluid temperature as a function of time, $T_{final}$ is the final temperature (80°C), $T_{initial}$ is the initial fluid temperature (15°C), e is Euler's number, t is time, and $\tau$ is the time constant. Equation 8 can also be used to predict that our experimental fluid heating time constant, bold red lines in Figure 5, is approximately 75 s.

Figure 9 shows the experimental inlet fluid temperature along with predicted charge and discharge response given a Heaviside step input for each module. Clearly, the time constants of the tested modules are low compared to that of the incoming fluid (75s), thus, it seems likely that the higher material FOM and improved time constant of NiTi was suppressed by the much longer fluid time constant. In effect, there is little perceived peak power benefit to using high-conductivity, high-FOM NiTi in instances, such as the one tested, where the external stimulus is much slower than the time constant of the module. This is confirmed by our peak power results in Figure 6. In this case, the thermal conductivity is of less importance and materials can be selected based more heavily on thermal capacitance, availability, and/or cost, for example. Nonetheless, in instances where the thermal driving force resembles a Heaviside step function or high frequency heat source, as would be the case in near-junction electronic cooling applications or the sudden opening of a fluid reservoir, the higher thermal conductivity of NiTi compared to 1-octadecanol would be expected to provide a lower time constant while maintaining high thermal capacity.

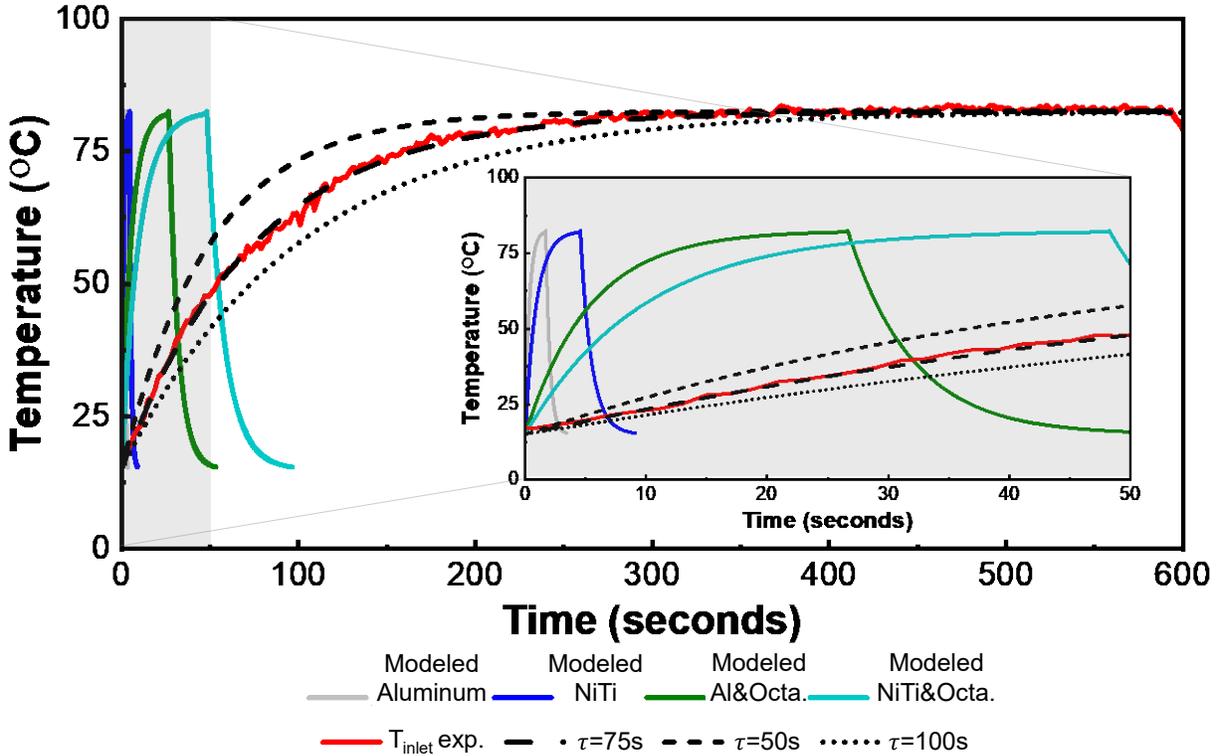

**Figure 9: Predicted (modeled) thermal response (Equation 7) for the four tested modules based on a heat transfer coefficient of 3950 Wm⁻²K⁻¹ and a Heaviside step function (instantaneous thermal boundary condition of 80°C) at t=0. T_inlet is the experimental inlet fluid temperature from Figure 5.**

Legend:
Modeled Aluminum, Modeled NiTi, Modeled Al&Octa., Modeled NiTi&Octa.
$T_{inlet}$ exp., $\tau=75s$, $\tau=50s$, $\tau=100s$

## 7. Conclusions

This paper reports the conceptualization, fabrication, and characterization of proof-of-concept solid-state nickel titanium thermal energy storage modules that store heat from, and reject heat to, water in a high power electronic cooling application. A commercially-available $Ni_{50.28}Ti_{49.36}$ material was characterized using differential scanning calorimetry and Xenon-flash diffusivity to demonstrate the high volumetric latent heat (183 kJm⁻³) and superior thermal conductivity (12.92 Wm⁻¹K⁻¹) over standard organic PCMs such as 1-octadecanol (183 kJm⁻³ and 0.15 Wm⁻¹K⁻¹). These exceptional transport properties, combined with the fact that nickel titanium does not melt, enabled the elimination of fin and encapsulant structures that have historically reduced PCM weight and volume in engineered modules. In the past, these design requirements have led to energy storage modules with PCM to total weight ratios of only 0.024 to 0.64 and volumetric power densities of at most 0.58 Wcm⁻³. Instead, we were able to fabricate modules that were nearly 100% PCM, enabling high power density operation up to 0.848 Wcm⁻³ as was the case in a composite NiTi/1-Octadecanol module. Furthermore, the high thermal conductivity of NiTi provides the additional benefit of smaller time constants (for a fixed or higher thermal capacity) and a clear path towards high- peak power and fast-transient operation. A common material figure-of-merit was shown to have limited applicability to system-level experimental performance due to discrepancies between realistic system forcing function and material time constants. Instead, lumped capacitance modeling proved more useful for understanding thermal performance. This indicates that careful consideration should be given to system forcing function time constants and anticipated TES time constants when developing prototypes and systems. Furthermore, an opportunity exists to improve upon existing predictive models and to formulate application-based FOMs that account for time constant by considering the relationship between module time constant and external forcing functions. More importantly, these pioneering results introduce metallic solid-state phase change materials to the thermal energy storage lexicon and offer a promising outlook on future developments involving high-capacity and high-power thermal energy storage.

## Supplementary Material

See supplementary material for information on the NiTi elemental analysis, NiTi material characterization, detailed uncertainty analysis, and a description of the system and module thermal time constants

## Author Information

*E-mail: darin.j.sharar.civ@mail.mil

## Acknowledgement

N/A

## Funding

This research did not receive any specific grant from funding agencies in the public, commercial, or not-for-profit sectors.

## Notes

The authors declare no competing financial interest.

## References

[1]  I. Mudawar, "Assessment of high-heat-flux thermal management schemes," IEEE Trans. Compon. Packag. Tech., vol. 24, no. 2, pp. 122-141, 2001.

[2]  H. de Bock, D. Huitink, P. Shamberger, J. Lundh, S. Choi, N. Niedbalski and L. Boteler, "A system to package perspective on transinet thermal management of electronics," Journal of Electronic Packaging, vol. 142, no. 4, 2020.

[3]   N. Jankowski and F. McCluskey, "A review of phase change materials for vehicle component thermal buffering," Applied Energy, vol. 113, pp. 1525-1561, 2014.

[4]   S. Monda, "Phase change materials for smart textiles - An overview," Applied Thermal Engineering, vol. 28, pp. 1536-1550, 2008.

[5]   F. Kuznik, D. David, K. Johannes and J. Roux, "A review on phase change materials integrated in building walls," Renewable Sustainable Energy, vol. 15, pp. 379-391, 2011.

[6]   A. de Gracia and L. Cabeza, "Phase change materials and thermal energy storage for buildings," Energy Build, vol. 103, pp. 414-419, 2015.

[7]   U. Pelay, L. Luo, Y. Fan, D. Stitou and M. Rood, "Thermal energy storage systems for concentrated solar power plants," Renew. and Sus. Energy Rev., vol. 79, pp. 82-100, 2017.

[8]   I. Sarbu and C. Sebarchievici, "A comprehensive review of thermal energy storage," Sustainability, vol. 10, p. 191, 2018.

[9]   H. Nazir, M. Batool, F. Bolivar Osorio, M. Isaza-Ruiz, X. Xu, K. Vignarooban, P. Phelan, Inamuddin and A. Kannan, "Recent developments in phase change materials for energy storage applications: A review," International Journal of Heat and Mass Transfer, vol. 129, pp. 491-523, 2019.

[10]  M. Yamaha and S. Misaki, "The evaluation of peak shaving by a thermal storage system using phase-change materials in air distribution systems," HVAC&R Res, vol. 12, pp. 861-869, 2006.

[11]  L. Cabeza, A. Castell, C. Barreneche, A. De Gracia and A. Fernandez, "Materials used as PCM in thermal energy storage in buildings: a review," Renew. Sustain. Energy Rev., vol. 15, pp. 1675-1695, 2011.

[12]  P. Shamberger and T. Fisher, "Cooling power and characteristic times of composite heatsinks and insulants," International Journal of Heat and Mass Transfer, vol. 117, pp. 1205-1215, 2018.

[13]  A. Hoe, M. Deckard, A. Tamraparni, A. Elwany, J. Felts and P. Shamberger, "Conductive heat transfer in lamellar phase change material composites," Applied Thermal Engineering, vol. 178, 2020.

[14]  M. Medrano, M. Yilmaz, M. Nogues, I. Martorell, J. Roca and L. Cabeza, "Experimental evaluation of commercial heat exchangers for use as PCM thermal storage systems," Applied Energy, vol. 86, no. 10, pp. 2047-2055, 2009.

[15]  J. Yang, L. Yang, C. Xu and X. Du, "Experimental study on enhancement of thermal energy storage with phase-change material," Applied Energy, vol. 169, pp. 164-176, 2016.

[16] A. Mallow, O. Abdelaziz and S. Graham, "Thermal charging study of compressed expanded natural graphite/phase change material composites," Carbon, pp. 495-504, 2016.

[17] X. Huang, G. Alva, L. Liu and G. Fang, "Microstructure and thermal properties of cetyl alcohol/high density polyethylene composite phase change materials with carbon fiber as shape-stabilized thermal storage materials," Applied Energy, vol. 200, pp. 19-27, 2017.

[18] Y. Tang, Y. Lin, Y. Jia and G. Fang, "Improved thermal properties of stearyl alcohol/high density polyethylene/expanded graphite composite phase change materials for building thermal energy storage," Energy Build., vol. 153, pp. 41-49, 2017.

[19] X. Wang, X. Cheng, Y. Li, G. Li and J. Xu, "Self-assembly of three-dimensional 1-octadecanol/graphene thermal storage materials," Sol. Energy, vol. 179, pp. 128-134, 2019.

[20] L. Chen, Z. Wub and J. Liu, "Carbon nanotube grafted with polyalcohol and its influence on the thermal conductivity of phase change material," Energy Convers. Manag., vol. 83, pp. 325-329, 2014.

[21] A. Al-Ahmed, A. Sari, M. Mazumder, G. Hekimoglu, F. Al-Sulaiman and Inamuddin, "Thermal energy storage and thermal conductivity properties of Octadecanol-MWCNT composite PCMs as promising organic heat storage materials," Nature, Scientific Reports, p. 15, 2020.

[22] H. Moon, N. Miljkovic and W. King, "High power density thermal energy storage using additively manufactured heat exchangers and phase change material," International Journal of Heat and Mass Transfer, vol. 153, p. 11, 2020.

[23] A. Iradukunda, A. Vargas, D. Huitink and D. Lohan, "Transient thermal performance using phase change material integrated topology optimized heat sinks," Applied Thermal Engineering, vol. 179, p. 8, 2020.

[24] D. Sharar, B. Donovan, R. Warzoha, A. Wilson, A. Leff and B. Hanrahan, "Solid-state thermal energy storage using reversible martensitic transformations," Applied Physics Letters, vol. 114, p. 6, 2019.

[25] D. Sharar, A. Wilson, A. Leff, A. Smith, K. Atli, A. Elwany, R. Arroyave and I. Karaman, "Additively manufacturing nitinol as a solid-state phase change material," ITHERM, p. 7, 2020.

[26] R. Warzoha, N. Vu, B. Donovan, E. Cimpoiasu, D. Sharar, A. Leff, A. Wilson and A. Smith, "Grain growth-induced thermal property enhancement of NiTi shape memory alloys for elastocaloric refrigeration and thermal energy storage systems," Int. J. of Heat and Mass Trans., vol. 154, p. 119760, 2020.

[27] G. Rondelli, "Corrosion resistance tests on NiTi shape memory alloy," Biomaterials, vol. 17, pp. 2003-2008, 1996.

[28] L. Casalena, A. Bucsek, D. Pagan, G. Hommer, G. Bigelow, M. Obstalecki, R. Noebe, M. Mills and A. Stebner, "Structure-property relationships of a high strength superelastic NiTi-Hf alloy," Adv. Eng. Mater., vol. 20, p. 1800046, 2018.

[29] H. Hou, E. Simsek, D. Stasak, N. Hasan, S. Qian, R. Ott, J. Cui and I. Takeuchi, "Elastocaloric cooling of additive manufactured shape memory alloys with large latent heat," J. Phys. D: Appl. Phys., vol. 50, p. 404001, 2017.

[30] J. Tang, M. Yang, F. Yu, X. Chen, L. Tan and G. Wang, "1-Octadecanol hierarchical porous polymer composite as a novel shape-stability phase change material for latent heat thermal energy storage," Applied Energy, vol. 187, pp. 514-522, 2017.

[31] P. Shamberger and N. Bruno, "Review of metallic phase change materials for high heat flux transient thermal management applications," Applied Energy, vol. 258, 2020.

[32] F. Trausel, A. de Jong and R. Cuypers, "A review on the properties of salt hydrates for thermochemical storage," Energy Procedia, vol. 48, pp. 447-452, 2014.

[33] G. Parravicini and A. Stella, "Extreme undercooling (down to 90K) of liquid metal nanoparticles," Applied Physics Letters, vol. 89, p. 5, 2006.

[34] H. Ryu, S. Woo, B. Shin and S. Kim, "Prevention of supercooling and stabilization of inorganic salt hydrates as latent heat storage materials," Sol. Energy Mater. Sol. Cell, vol. 27, no. 2, pp. 161-172, 1992.

[35] A. Fernandez, C. Barreneche, M. Belusko, M. Segarra, F. Bruno and L. Cabeza, "Considerations for the use of metal alloys as phase change materials for high temperature applications," Solar Energy Materials and Solar Cells, vol. 171, pp. 275-281, 2017.

[36] M. Kubota, E. Ona, H. Watanabe, H. Matsuda, H. Hidaka and H. Kakiuchi, "Studies on phase change characteristics of binary mixtures of erythritol and MgCl-6H2O," J. Chem. Eng. Jpn., vol. 40, no. 1, pp. 80-84, 2007.

[37] G. Lane, Solar heat storage: latent heat materials. Background and scientific principles, vol.1, Boca Raton, FL: CRC Press, 1983.

[38] M. Fazilati and A. Lemrajabi, "Phase change material for enhancing solar water heater, an experimental approach," Energy conversion and Management, vol. 71, pp. 138-145, 2013.

[39] R. Zarnetta, R. Takahashi, M. Young, A. Savan, Y. Furuya, S. Thienhaus, B. MaaB, M. Rahim, J. Frenzel, H. Brunken, Y. Chu, V. Srivastava, R. James, I. Takeuchi, G. Eggeler and A. Ludwig, "Identification of quaternary shape memory alloys with near-zero thermal hysteresis and unprecendented functional stability," Advanced Functional Materials, vol. 20, pp. 1917-1923, 2010.

[40] X. Liu, M. Hu, C. Caneau, R. Bhat and C. Zah, "Thermal management strategies for high power semiconductor pump lasers," IEEE Trans Compon Packag Tech, vol. 29, pp. 268-276, 2006.

[41] C. Chluba, W. Ge, R. de Miranda, J. Strobel, L. Kienle, E. Quandt and M. Wuttig, "Ultra-low fatigue shape memory alloy films," Science, vol. 348, pp. 1004-1007, 2015.

[42] J. Frenzel, A. Wieczorek, I. Opahle, B. MaaB, R. Drautz and G. Eggeler, "On the effect of alloy composition on martensite start temperatures and latent heats in Ni-Ti-based shape memory alloys," Acta Materialia, vol. 90, pp. 213-231, 2015.

[43] L. Casalena, G. Bigelow, Y. Gao, O. Benafan, R. Noebe, Y. Wang and M. Mills, "Mechanical behavior and microstructural analysis of NiTi-40Au shape memory alloys exhibiting work output above 400C," Intermetallics, vol. 86, pp. 33-44, 2017.

[44] A. Bucsek, G. Hudish, G. Bigelow, R. Noebe and A. Stebner, "Composition, compatibility, and the functional performances of ternary NiTiX high-temperature shape memory alloys," Shape Mem. Superelasticity, vol. 2, pp. 62-79, 2016.

[45] K. Mohanchandra, D. Shin and G. Carman, "Deposition and characterization of Ti-Ni-Pd and Ti-Ni-Pt shape memory alloy thin films," Smart Mater. Struct., vol. 14, pp. 312-316, 2005.

[46] L. Casalena, D. Coughlin and F. Yang, "Transformation and deformation characterization of NiTiHf and NiTiAu high temperature shape memory alloys," Microsc. Micoanal., vol. 21, pp. 607-608, 2003.

[47] N. Zarubova and V. Novak, "Phase stability of CuAlMn shape memory alloys," Materials Science and Engineering: A, vol. 378, no. 1-2, pp. 216-221, 2004.

[48] B. Xu, P. Li and C. Chan, "Extending the validity of lumped capacitance method for large Biot number in thermal storage application," Solar Energy, vol. 86, pp. 1709-1724, 2012.

[49] T. Lu, "Thermal management of high power electronics with phase change cooling," International Journal of Heat and Mass Transfer, vol. 43, pp. 2245-2256, 2000.

[50] A. Smith and H. Nochetto, "Laminar thermally developing flow in rectangular channels and parallel plates: uniform heat flux," Heat & Mass Transfer, vol. 50, pp. 1627-1637, 2014.

**SUPPLEMENTARY INFORMATION**

**Elemental Analysis**

An elemental analysis was performed by EdgeTech and came listed by chemical composition and % weight on a 'Certificate of Analysis'. The % weight values for the three materials tested are transcribed in Table S1-S3, along with calculated atomic % values. The atomic %s reported in the main text, $Ni_{50.28}Ti_{49.36}$, $Ni_{50.31}Ti_{49.32}$, and $Ni_{50.41}Ti_{49.20}$, were calculated based on the reported % weight from EdgeTech and known atomic mass of constituent elements, as shown below.

**Table S1: % weight, elemental composition, elemental atomic mass, and atomic % calculation of the $Ni_{50.28}Ti_{49.36}$ material**

| Quantity | Ni | Ti | Cr | Cu | Fe | Nb | O | N | C | Co | H |
|---|---|---|---|---|---|---|---|---|---|---|---|
| % weight | 55.48 | 44.414 | 0.005 | 0.005 | 0.012 | 0.005 | 0.037 | 0.001 | 0.035 | 0.005 | 0.001 |
| Atomic mass | 58.69 | 47.87 | 51.99 | 63.55 | 55.85 | 92.91 | 15.99 | 14.01 | 12.01 | 58.93 | 1.01 |
| atomic % | 50.28 | 49.35 | 0.00512 | 0.00419 | 0.01143 | 0.00286 | 0.12301 | 0.0038 | 0.15501 | 0.00451 | 0.05278 |

**Table S2: % weight, elemental composition, elemental atomic mass, and atomic % calculation of the $Ni_{50.31}Ti_{49.32}$ material**

| Quantity | Ni | Ti | Cr | Cu | Fe | Nb | O | N | C | Co | H |
|---|---|---|---|---|---|---|---|---|---|---|---|
| % weight | 55.51 | 44.38 | 0.005 | 0.005 | 0.012 | 0.005 | 0.037 | 0.001 | 0.035 | 0.005 | 0.001 |
| Atomic mass | 58.69 | 47.87 | 51.99 | 63.55 | 55.85 | 92.91 | 15.99 | 14.01 | 12.01 | 58.93 | 1.01 |
| atomic % | 50.31 | 49.32 | 0.00512 | 0.00419 | 0.01143 | 0.00286 | 0.12301 | 0.0038 | 0.15501 | 0.00451 | 0.05278 |

**Table S3: % weight, elemental composition, elemental atomic mass, and atomic % calculation of the $Ni_{50.41}Ti_{49.20}$ material**

| Quantity | Ni | Ti | Cr | Cu | Fe | Nb | O | N | C | Co | H |
|---|---|---|---|---|---|---|---|---|---|---|---|
| % weight | 55.62 | 44.271 | 0.005 | 0.005 | 0.012 | 0.005 | 0.04 | 0.001 | 0.035 | 0.005 | 0.001 |
| Atomic mass | 58.69 | 47.87 | 51.99 | 63.55 | 55.85 | 92.91 | 15.99 | 14.01 | 12.01 | 58.93 | 1.01 |
| atomic % | 50.41 | 49.20 | 0.00512 | 0.00419 | 0.01143 | 0.00286 | 0.13299 | 0.0038 | 0.15501 | 0.00451 | 0.05278 |

## Additional NiTi Material Characterization

DSC results for the as-received and heat treated $Ni_{50.28}Ti_{49.36}$, $Ni_{50.31}Ti_{49.32}$, and $Ni_{50.41}Ti_{49.20}$ materials are shown in Figure S1 and summarized in the adjacent table. As expected, higher Ni content resulted in lower transformation temperature [42], varying from Austenite peak ($A_p$) temperatures of 76°C for $Ni_{50.28}Ti_{49.36}$ to 13°C for $Ni_{50.41}Ti_{49.20}$. Both $Ni_{50.28}Ti_{49.36}$ and $Ni_{50.31}Ti_{49.32}$ provide full transformations in the desired temperature range of 15°C and 80 °C. However, the $Ni_{50.41}Ti_{49.20}$ material did not provide a reverse transformation within the same range. The latent heat showed an inverse relationship with Ni content, meaning higher transformation temperature corresponded with higher latent heat. Furthermore, the solution anneal process proved useful for increasing the measured latent heat with no significant impact on transformation temperature. In the case of $Ni_{50.28}Ti_{49.36}$, the latent heat was shown to increase by 230%, from 12 to 28 $Jg^{-1}$.

The thermal diffusivity of the as-received and solution heat treated NiTi materials were measured using a TA Instruments DXF 200 high-speed Xenon-pulse delivery source and solid-state PIN detector. The thermal conductivity, calculated from the diffusivity assuming a density of 6450 $kgm^{-3}$ and specific heat of 469 $Jkg^{-1}K^{-1}$ [26], for all NiTi compositions at a temperature of $0°C$ (Martensite phase) and $100°C$ (Austenite phase) are listed in Figure 3aS1. Similar to the latent heat trends, there appears to be a positive relationship between transformation temperature and thermal conductivity, with $Ni_{50.28}Ti_{49.36}$ outperforming the others for the full range of temperatures and processing conditions tested.

Both $Ni_{50.28}Ti_{49.36}$ and $Ni_{50.31}Ti_{49.32}$ could be used effectively as PCMs for the current application, but the solution heat treated $Ni_{50.28}Ti_{49.36}$ offers the highest latent heat (28.30 vs 20.51 $Jg^{-1}$) and highest thermal conductivity (12.92 vs 12.22 $Wm^{-1}K^{-1}$) of the two. Based on these results, solution heat treated $Ni_{50.28}Ti_{49.36}$ was chosen as the solid-state material to be used in the thermal energy storage unit in this study. It's worth noting that $Ni_{50.41}Ti_{49.20}$ would not be an appropriate choice because it lacks a full transformation in the specific temperature range. Nonetheless, this characterization demonstrates a key benefit of the class of material: the ability to tailor transformation temperature, latent heat, and thermal conductivity by altering chemical composition and heat treatment.

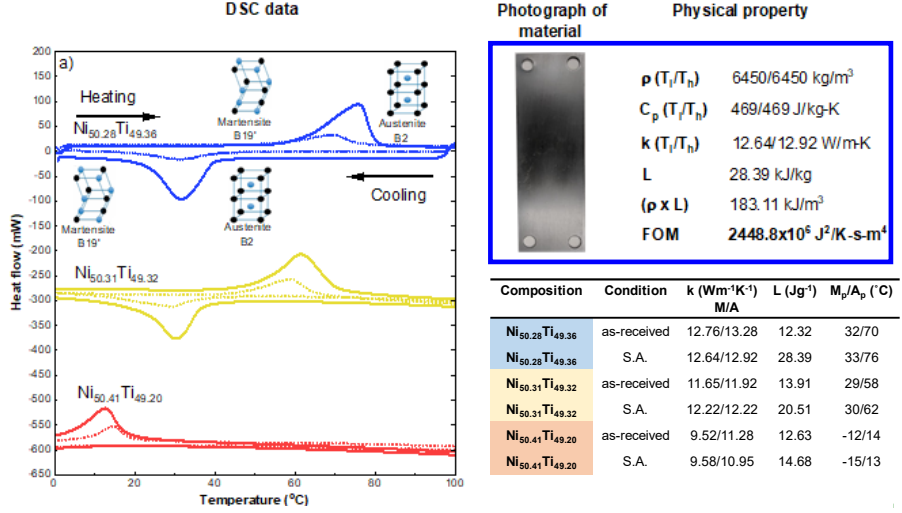

**Figure S1: DSC data, photographs, and physical properties of the materials used in the current study and calculated PCM Figure of Merit: a) 1-Octadecanol, b) aluminum 6061, and c) Ni$_{50.28}$Ti (dashed line is the as-received material and solid line is the heat treated material)**

Physical property box:

| Property | Value |
|---|---|
| $\rho$ (T$_l$/T$_h$) | 6450/6450 kg/m³ |
| C$_p$ (T$_l$/T$_h$) | 469/469 J/kg-K |
| k (T$_l$/T$_h$) | 12.64/12.92 W/m·K |
| L | 28.39 kJ/kg |
| ($\rho$ x L) | 183.11 kJ/m³ |
| FOM | 2448.8x10⁶ J²/K·s·m⁴ |

| Composition | Condition | k (Wm⁻¹K⁻¹) M/A | L (Jg⁻¹) | M$_d$/A$_p$ (°C) |
|---|---|---|---|---|
| Ni$_{50.28}$Ti$_{49.36}$ | as-received | 12.76/13.28 | 12.32 | 32/70 |
| Ni$_{50.28}$Ti$_{49.32}$ | S.A. | 12.64/12.92 | 28.39 | 33/76 |
| Ni$_{50.31}$Ti$_{49.32}$ | as-received | 11.65/11.92 | 13.91 | 29/58 |
| Ni$_{50.31}$Ti$_{49.32}$ | S.A. | 12.22/12.22 | 20.51 | 30/62 |
| Ni$_{50.41}$Ti$_{49.20}$ | as-received | 9.52/11.28 | 12.63 | -12/14 |
| Ni$_{50.41}$Ti$_{49.20}$ | S.A. | 9.58/10.95 | 14.68 | -15/13 |

## Uncertainty Analysis

As shown in Equation 4 in the main text, the error in the reported instantaneous power is a function of the mass flowrate and the difference between the inlet and outlet temperature, $\Delta T$:

$$\frac{\delta q_{instantaneous}}{q_{instantaneous}} = \sqrt{\left(\frac{\delta \dot{m}}{\dot{m}}\right)^2 + \left(\frac{\delta \Delta T}{\Delta T}\right)^2} \qquad \text{(S1)}$$

The error in the experimental energy absorbed by the module, is a function of the error in the power, $\bar{q}$, the error in the estimated energy absorbed by the tubing, and the error in the parasitic heat loss to the environment:

$$\frac{\delta E}{E} = \sqrt{\left(\frac{\delta \bar{q}}{\bar{q}}\right)^2 + \left(\frac{\delta E_{tubing}}{E_{tubing}}\right)^2 + \left(\frac{\delta E_{loss}}{E_{loss}}\right)^2} \qquad \text{(S2)}$$

where the uncertainty in the Energy loss is:

$$\frac{\delta E_{loss}}{E_{loss}} = \sqrt{\left(\frac{\delta \dot{m}}{\dot{m}}\right)^2 + \left(\frac{\delta \Delta T}{\Delta T}\right)^2} \qquad \text{(S3)}$$

The uncertainty in the energy absorbed by the ancillary stainless steel tubing is:

$$\frac{\delta E_{tubing}}{E_{tubing}} = \sqrt{\left(\frac{\delta m}{m}\right)^2 + \left(\frac{\delta \Delta T}{\Delta T}\right)^2} \qquad \text{(S4)}$$

And the uncertainty in the average power is:

$$\frac{\delta \bar{q}}{\bar{q}} = \sqrt{\sum_{t_{start}}^{t_{finish}} \left(\frac{\delta q_{instantaneous}}{q_{instantaneous}}\right)^2} \tag{S5}$$

Table S4 is a summary of uncertainties for measured and calculated experimental parameters.

**Table S4: Summary of uncertainty associated with the primary parameters of interest**

| Measured/Calculated parameter | Uncertainty |
|---|---|
| mass flowrate ($\dot{m}$) | $\pm$ 1.5% (kg/s) |
| mass of stainless steel tubing ($m$) | $\pm$ 0.01 (g) |
| Temperature ($T$) | $\pm$ 0.5 (K) |
| $E_{tubing}$ (assuming 65K $\Delta T$) | $\pm$ 0.77% (J) |
| $E_{loss}$ (at high temperature) | $\pm$ 6.22% (J) |
| Temperature difference ($\Delta T$) | 4.23% < $\delta$ < 228% (K) |
| $q_{instantaneous}$ | 4.49% < $\delta$ < 228% (W) |

## Detailed Calculation Methods and Thermal Capacity Results

Table S5 lists the salient physical features of each fabricated thermal energy storage module (size, weight, heat transfer area, etc.), including estimated energy storage capacities for a temperature change of 65˚C, from 15˚C to 80˚C, to match the experiments. The sensible heat capacity of the modules was estimated based on the following equation:

$$q_{sensible} = \left[mC_{p\,low\,temp}(T_{transformation} - T_{initial}) + mC_{p\,high\,temp}(T_{final} - T_{transformation})\right]_{SL} + \left[mC_p(T_{final} - T_{initial})\right]_{SS} + \left[mC_p(T_{final} - T_{initial})\right]_{Al} \tag{S6}$$

where $q_{sensible}$ is the total sensible heat capacity (J), $m$ is the mass of material (kg), $C_p$ is the specific heat (Jkg$^{-1}$K$^{-1}$), and $T$ is the start, transformation, or final temperature of the material.

The latent heat capacity of the modules was estimated based on the following equation:

$$q_{latent} = [mL]_{SL} + [mL]_{SS} \tag{S7}$$

where $q_{latent}$ (J) is the total latent heat of the module, $m$ is the mass (g), and $L$ is the latent heat (Jg$^{-1}$). The total sensible and latent heat for a given module is simply the sum of Equation S6 and S7. Finally, the mass specific (kJkg$^{-1}$) and volume specific (kJcm$^{-3}$) energy densities were estimated by dividing the module total energy storage by the module mass and volume, respectively.

**Table S5: 3D diagrams, fluidic paths, and salient physical and thermal characteristics of TES modules explored (grey area indicates values calculated based on experimental results in Figures 5 & 6)**

| 3D exploded diagram and fluid flow paths | | | | |
|---|---|---|---|---|
| **Description** | Al | **Al & 1-Octadecanol** | **NiTi** | **NiTi & 1-Octadecanol** |
| **Total Volume (cm³)** | 187.5 | 187.5 | 187.5 | 187.5 |
| **Total mass with PCM (g)** | 202.2 | 235.4 | 537.7 | 570.9 |
| **Heat transfer area (cm²)** | 1342.2 | 745.7 | 1342.2 | 745.7 |
| **Mass PCM (g) SL/SS** | 0/0 | 33.2/0 | 0/537.7 | 33.2/537.7 |
| **Volume PCM (cm³) SL/SS** | 0/0 | 41/0 | 0/74.6 | 41/74.6 |
| **PCM sensible heat capacity* (kJ) SL/SS** | 0/0 | 5.6/0 | 0/16.4 | 5.6/16.4 |
| **Module sensible heat capacity* (kJ)** | 11.83 | 17.5 | 16.4 | 22.0 |
| **Latent energy storage (kJ) SL/SS** | 0/0 | 7.5/0 | 0/15.1 | 7.5/15.1 |
| **Module sensible & latent storage (kJ)** | 11.83 | 24.9 | 31.5 | 44.5 |
| **Mass specific energy storage* (kJ/kg)** | 58.5 | 105.7 | 58.6 | 78.0 |
| **Volume specific energy storage* (kJ/cm³)** | 0.063 | 0.133 | 0.168 | 0.238 |
| **Measured sensible & latent storage (kJ)** | 12.1 | 23.8 | 30.9 | 41.2 |
| **Mass specific energy storage (kJ/kg)** | 59.8 | 101.1 | 57.5 | 72.2 |
| **Volume specific energy storage (kJ/cm³)** | 0.064 | 0.127 | 0.1648 | 0.219 |

## Time constant determination

The time constant for the four energy storage modules can be approximated using the following general equation:

$$\tau = \frac{\rho C_p V + \frac{L}{T_{final} - T_{initial}}}{h_{eff} A_s} \qquad (S8)$$

where $\rho$ is the density, $C_p$ is the specific heat, $V$ is the volume, $h_L$ is the latent heat, $h_{eff}$ is the effective heat transfer coefficient, and $A_s$ is the convective heat transfer area. The latent heat was distributed over the full temperature range according to the expression, $\frac{L}{T_{final} - T_{initial}}$, to reduce numerical complexity. For a calculated heat transfer coefficient of 3950 Wm⁻¹K⁻¹, the Biot number for our modules ranges from 0.01 to 10. Therefore, conductive thermal resistance is not negligible for all scenarios, and a standard lumped capacitance method is not strictly appropriate. Instead, a modified $h_{eff}$ may be used to account for the non-negligible conductive resistance; for parallel plates, Xu et al. [48] defined the effective heat transfer coefficient to be:

$$h_{eff} = \frac{1}{\frac{1}{h} + \frac{(l/2)}{3k}} \qquad (S9)$$

where h is the actual heat transfer coefficient, l is the plate thickness, and k is the plate thermal conductivity. Using Equations S8 and S9, the time constant for the four modules were reported in the main body of the text.

Supporting calculations were necessary to determine the thermal time constants of the energy storage modules, namely the heat transfer coefficient (defined as h in Equation S9). The full heat transfer coefficient calculation was not deemed pertinent to the text, nor was it unique, and is included here for reference. The correlation by Smith and Nochetta [50] was used to predict local h values along the channel length assume an infinite aspect ratio. Nusselt number can be calculated from their paper:

$$Nu_{D_h, x/m} = Cx^* \qquad (S10)$$

and

$$Nu_{D_h} \cong \left[ \left( Nu_{D_h, x^* \to 0} \right)^N + \left( Nu_{D_h, x^* \to \infty} \right)^N \right]^{1/N} \qquad (S11)$$

and using local Nusselt number curve fit parameters from Table 4 (of their paper):

$Nu_{fd} = 8.235$

$C = 1.521$

$m = -0.3314$

$N = 4.761$

From these parameters, we were able to predict the heat transfer coefficient as a function of channel length. As shown in Figure S2, the heat transfer coefficient varied between 5600 and of 3900 $Wm^{-2}K^{-1}$. Please note, the x-axis is from 0 to 0.154m (0 to 6 inches) and is modeled here as flow through a parallel plate with uniform channel geometry. The heat transfer coefficient is higher at the inlet (x=0 m) than the outlet (0.154 m) due to developing flow effects, as would be expected. The average value for Figure S2 is 3950 $Wm^{-2}K^{-1}$, as referenced in the main text.

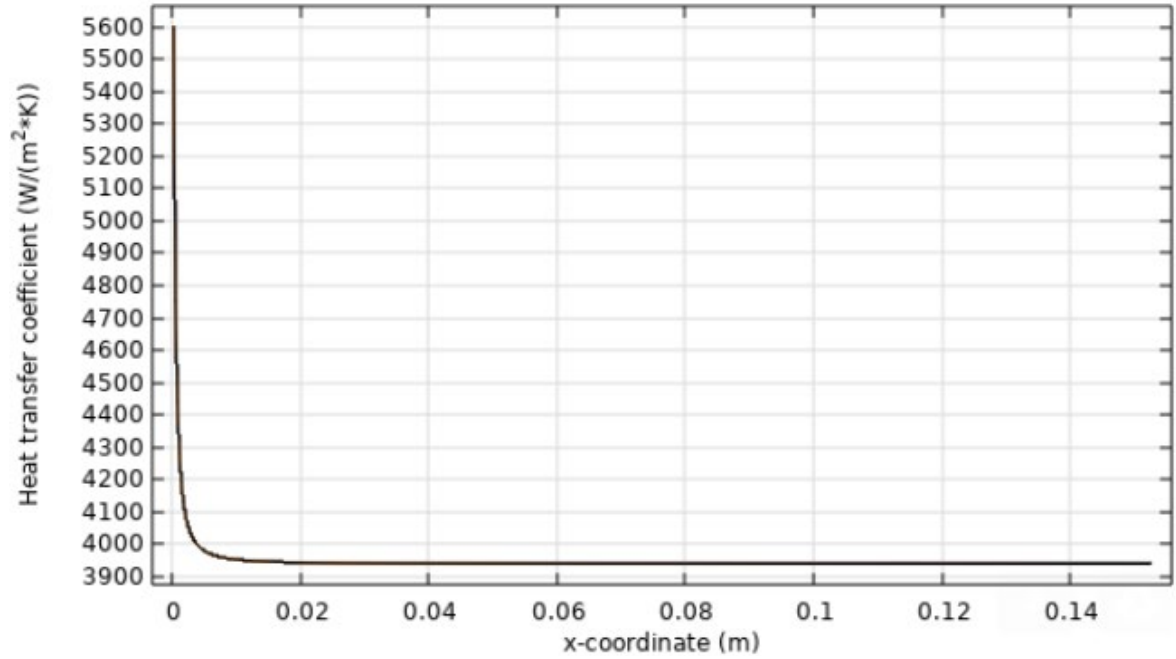

**Figure S2: Calculated heat transfer coefficient as a function of distance along the parallel plate flow channel based on the correlation from Smith and Nochetta [43]. The developing flow correlation varies from 5600 at the inlet to 3900 $Wm^{-1}K^{-1}$ at the outlet, with an average value of 3950 $Wm^{-1}K^{-1}$. The average value is reported in the main manuscript and was used to calculate the module thermal time constants.**